\documentstyle[amssymb,12pt,psfig]{article}

\input{tcilatex}

\begin{document}

\title{Anomalous behavior of \\
ideal Fermi gas below two dimensions}
\author{M. Grether$^{1}$, M. de Llano$^{2}$ and M.A. Sol\'{\i}s$^{3}$ \\
$^{1}$Facultad de Ciencias, UNAM, Apdo. Postal 70-542,\\
04510 M\'{e}xico, DF, Mexico \\
$^{2}$Instituto de Investigaciones en Materiales, UNAM, \\
Apdo. Postal 70-360, 04510 M\'{e}xico, DF, Mexico \\
$^{3}$Department of Physics, Washington University \\
St. Louis, Missouri 63130, USA and \\
Instituto de F\'{\i}sica, UNAM, Apdo. Postal 20-364,\\
01000 M\'{e}xico, DF, Mexico}
\maketitle

\begin{abstract}
Normal behavior of the thermodynamic properties of a Fermi gas in $d>2$
dimensions, integer or not, means monotonically increasing or decreasing of
its specific heat, chemical potential or isothermal sound velocity, all as
functions of temperature. However, for $0<d<2$ dimensions these properties
develop a ``hump'' (or ``trough'') which increases (or deepens) as $%
d\rightarrow 0$. Though not the phase transition signaled by the sharp
features (``cusp'' or ``jump'') in those properties for the ideal Bose gas
in $d>2$ (known as the Bose-Einstein condensation), it is nevertheless an
intriguing structural anomaly which we exhibit in detail.


\end{abstract}

\section{Introduction}

The familiar ideal Fermi gas is revisited for any positive space dimension, $%
d>0$. Ideal Fermi gases (IFG) have been discussed thoroughly in general by
several authors \cite{Path}-\cite{tsallis} and a detailed study of the
quantum behavior in any dimension at sufficiently low temperatures in these
systems have also gained interest as possible precursors of a paired-fermion
condensate at lower temperatures, these have been studied experimentally in
ultra-cold fermionic clouds, e.g., with $_{19}^{40}$K neutral atoms in
opto-magnetic traps \cite{MarcoJin}-\cite{Holland}. From the thermodynamic
grand potential we generalize the system to any dimension, we calculated the
thermodynamics properties and analyzed the results for the chemical
potential, the heat capacity and the isothermal velocity of sound where we
found that there are \textit{structure} if $d<2$ in contrast with the ideal
Boson gas where there is structure if $d>2$ \cite{Ziff, Agui}. In Sec.
2, we calculate the thermodynamic grand potential for the non-interacting
Fermi gas in $d$-dimensions, we find the thermodynamic properties of these
systems and deduce a generalized density of states (DOS). In Sec. 3, we
calculate the chemical potential $\mu (T),$ the internal energy and the
specific heat as function of absolute temperature $T$ \ for any positive
space dimension $d$ and analyzed the results obtained for $d<2$. In Sec. 4
we obtain the isothermal and adiabatic velocities of sound for these
systems. Sec. 5 contains our conclusions.

\section{Ideal Fermi gas in $d>0$ dimensions}

We consider an ideal quantum gas\ of \ fermions\ in $d>0$ dimensions of mass 
$m$ in vacuum with a quadratic dispersion relation. The Hamiltonian is then $%
H=\sum_{i=1}^{d}p_{i}^{2}{/2m}$, its eigenvalues are given by 
\begin{equation}
\varepsilon _{n}={\frac{2\pi ^{2}\hbar ^{2}}{mL^{2}}}\sum_{i=1}^{d}n_{i}^{2},
\label{ener}
\end{equation}
where $L$ is the size of the ``box'', $\mathbf{n}=(n_{1},n_{2},...,n_{d})$,
and where $n_{i}=0,\ \pm 1,\;\pm 2,\;...$. Since $k_{i}\equiv (2\pi /L)n_{i}$%
, {(\ref{ener}) can then be rewritten as } 
\begin{equation}
\varepsilon _{k_{i}}={\frac{\hbar ^{2}}{2m}}\sum_{i=1}^{d}k_{i}^{2}.
\end{equation}
The thermodynamical properties of this system follow from the thermodynamic
(or grand potential) $\Omega (T,L^{d},\mu )=U-TS+\mu N,$\ where $U$ is the
internal energy, $T$ the absolute temperature, $S$ the entropy, $\mu $ the
chemical potential, and $N$ the number of particles. We may write $\Omega
(T,L^{d},\mu )$ in generalized form as (see p. 134 of \cite{Path}) 
\begin{equation}
\Omega (T,L^{d},\mu )=-k_{B}T\sum_{{k_{i}}}\hbox{ln}[1+{\normalsize e}^{-%
{\normalsize \beta }({\normalsize \varepsilon }_{k_{i}}{\normalsize -\mu }%
)}],  \label{om}
\end{equation}
with $\beta \equiv 1/k_{B}T$, and $k_{B}$ is the Boltzmann constant. Using
the logarithm expansion ln$(1+x)=-\sum_{l=1}^{\infty }(-x)^{l}/l${\ valid
for }${x<1}$, {(\ref{om}) becomes} 
\begin{eqnarray}
\Omega (T,L^{d},\mu ) &=&k_{B}T\sum_{{k_{i}}}\sum_{{l}=1}^{\infty }{\frac{%
(-e^{-\beta (\varepsilon _{n}-\mu )})^{{l}}}{{l}}}  \nonumber \\
{} &=&{}k_{B}T\sum_{{l}=1}^{\infty }{\frac{(-e^{\beta \mu })^{{l}}}{{l}}}%
\sum_{{k_{i}}}e^{-\beta {l}\left[ (\hbar ^{2}/2m{)}\sum_{i=1}^{d}k_{i}^{2}%
\right] }.  \label{ome}
\end{eqnarray}

In the continuous limit where $\hbar ^{2}/mL^{2}\ll k_{B}T$, the summation
over $k_{i}$ can be approximated by integral, namely ${\ }\sum_{k_{i}}\ \ {%
\longrightarrow \ \ }(2 {\mathtt{s}}+1)(L/2\pi )^{d}\int d^{d}k_{i}$. Thus 
\begin{eqnarray}
\Omega (T,V,\mu ) &=&k_{B}T(2{\mathtt{s}}+1)(L/2\pi )^{d}\sum_{l=1}^{\infty }{%
\frac{(-e^{\beta \mu })^{{l}}}{{l}}}\int_{-\infty }^{\infty }dk_{1}\ e^{{%
-\beta }\mathit{l}(\hbar ^{2}/2m{)}k_{1}^{2}}\times   \nonumber \\
&&\int_{-\infty }^{\infty }dk_{2}\ e^{{-\beta }\mathit{l}(\hbar ^{2}/2m{)}%
k_{2}^{2}}\quad \cdots \quad \int_{-\infty }^{\infty }dk_{d}\ e^{{-\beta }%
\mathit{l}(\hbar ^{2}/2m{)}k_{d}^{2}},  \label{ome1}
\end{eqnarray}
with $\mathtt{s}$ the particle spin. Integrating, Eq. (\ref{ome1}) becomes 
\begin{equation}
\Omega =\left( 2\mathtt{s}+1\right) {\beta ^{-(d/2+1)}}\left( {\frac{mL^{2}}{%
2\pi \hbar ^{2}}}\right) ^{d/2}\sum_{l=1}^{\infty }{\frac{(-e^{\beta \mu })^{%
{l}}}{{l}^{d/2+1}}}.  \label{ome2}
\end{equation}
The infinite sum can be expressed in terms of the Fermi functions $f_{\sigma
}(z)$ (see Appendix D of Ref. \cite{Path}), 
\begin{equation}
f_{\sigma }(z)\equiv -\sum_{l=1}^{\infty }{\frac{(-z)^{{l}}}{{l}^{\sigma }}=%
\frac{1}{\Gamma (\sigma )}}\int_{0}^{\infty }dx{\frac{x^{\sigma -1}}{%
z^{-1}e^{x}+1},}  \label{fermi}
\end{equation}
so that 
\begin{equation}
{\Omega }=-\left( 2\mathtt{s}+1\right) \left( {\frac{mL^{2}}{2\pi \hbar ^{2}}%
}\right) ^{d/2}\frac{f_{d/2+1}(z)}{{\small \beta }^{d/2+1}}\equiv -A_{d}%
\frac{f_{d/2+1}(z)}{{\small \beta }^{d/2+1}},  \label{TGP}
\end{equation}
which defines $A_{d}$, and where $z\equiv e^{\beta \mu (T)}$.

From (\ref{TGP}) it is possible to find the thermodynamic properties of a
monatomic gas using the relation $d\Omega =-SdT-PdV-Nd\mu ,$ where $V\equiv
L^{d}$ is the system volume and $P$ its pressure. In this representation,
the grand potential $\Omega (T,V,\mu )$ is the fundamental relation leading
to all the thermodynamic variables of the system, namely, 
\begin{equation}
S=-\left( {\frac{\partial \Omega }{\partial T}}\right) _{V,\mu },\quad
P=-\left( {\frac{\partial \Omega }{\partial V}}\right) _{T,\mu }=-\frac{%
\Omega }{V}{\ \ \ \ }{\mbox{and}}{\ \ \ }N=-\left( {\frac{\partial \Omega }{%
\partial \mu }}\right) _{T,V}.  \label{est}
\end{equation}

\section{Thermodynamic variables}

\subsection{\protect\bigskip Chemical potential}

We determine the chemical potential $\mu (T)$ as a function of absolute
temperature $T$ \ for any positive space dimension $d>0$ from the number
equation which is given from (\ref{TGP}) and (\ref{est}),

\begin{equation}
N=\frac{A_{d}}{{\small \beta }^{d/2+1}}\left( \frac{\partial }{\partial \mu }%
f_{d/2+1}(z)\right) _{T,V}=A_{d}\frac{f_{{d/2}}(z)}{\beta ^{d/2}},  \label{N}
\end{equation}
which after substituting $A_{d}$ in (\ref{N}) and using (\ref{fermi}) becomes

\begin{equation}
N=\frac{\left( 2\mathtt{s}+1\right) }{\Gamma \left( d/2\right) }\left( {%
\frac{mL^{2}}{2\pi \hbar ^{2}}}\right) ^{d/2}\int_{0}^{\infty }d\varepsilon {%
\frac{\varepsilon ^{{d/2}-1}}{e^{\beta (\varepsilon -\mu )}+1}}.  \label{N1}
\end{equation}
Since $N=\int_{0}^{\infty }d\varepsilon {g(\varepsilon )n(\varepsilon )}$
where $\ {g(\varepsilon )}$ is the density of states (DOS) and $%
n(\varepsilon )\equiv \left[ e^{\beta (\varepsilon -\mu )}+1\right] ^{-1}$is
the Fermi-Dirac distribution, then 
\begin{equation}
g(\varepsilon )=\left( 2\mathtt{s}+1\right) \left( \frac{{mL}^{2}}{2\pi
\hbar ^{2}}\right) ^{d/2}\frac{\varepsilon ^{{d/2}-1}}{\Gamma \left(
d/2\right) }.  \label{dos}
\end{equation}
Since $\left[ e^{\beta \{\varepsilon -\mu (T)\}}+1\right] ^{-1}%
\smash {\
\mathop{\relbar\joinrel\longrightarrow}\limits_{T\to 0}\ \ }\theta \left(
E_{F}-\varepsilon \right) $, with\ $\theta (x)$ the unit step function, $\mu
\left( 0\right) \equiv $\ $E_{F}\equiv \hbar ^{2}k_{F}^{2}/2m$ the Fermi
energy, $k_{F}$ being the Fermi wavenumber, we see\ from (\ref{N1}) that 
\begin{equation}
n\equiv \frac{N}{L^{d}}\ {\smash
{\mathop{\relbar\joinrel\longrightarrow}\limits_{T\to 0}}\ \ }\frac{{2}%
\left( 2\mathtt{s}+1\right) }{d\Gamma (d/2)}\left( {\frac{m}{2\pi \hbar ^{2}}%
}\right) ^{d/2}E_{F}^{d/2},  \label{ro0}
\end{equation}
and we recover the expressions obtained in \cite{ariel, casas1} for the
fermion number density with $\mathtt{s}=1/2,$ i.e., 
\begin{equation}
n\equiv \frac{N}{L^{d}}=\frac{k_{F}^{d}}{2^{d-2}\pi ^{d/2}d\Gamma (d/2)},
\label{dend}
\end{equation}
which gives the familiar results $2k_{F}/\pi ,$ $k_{F}^{2}/2\pi $ and $%
k_{F}^{3}/3\pi ^{2}$ for $d=1,$ $2$ and $3$, respectively. Defining $%
E_{F}\equiv k_{B}T_{F},$ the chemical potential $\mu (T)$ is obtained from (%
\ref{N1}) and (\ref{ro0}) by solving numerically the following equation 
\begin{equation}
T^{d/2}\Gamma (d/2)f_{d/2}(e^{\beta \mu })=\left( 2/d\right) T_{F}^{\ d/2}.
\label{FFD0}
\end{equation}

For $d=2$, $f_{1}(z)$ from (\ref{fermi}) is the ordinary log function so
that (\ref{FFD0}) gives 
\begin{equation}
f_{1}(z)=\ln (1+z)=T_{F}{\large /}T,  \label{PolyLog}
\end{equation}
leading to the relatively well-known explicit \cite{mckel}{\Huge \ }formula 
\begin{equation}
\mu (T)/E_{F}=T/T_{F}\ln \left( e^{{T}_{F}/T}-1\right) \quad \smash
{\mathop{\relbar\joinrel\longrightarrow}\limits_{T\to 0}}\quad 1\,,
\label{muT}
\end{equation}
which is clearly not expandable in powers of $\ {T/T}_{F}$ as in the
so-called Sommerfeld expansion \cite{ash}. However, for $d\neq 2$, $\mu (T)$
is not an explicit, closed expression in $T/T_{F}$; numerical analysis is
required to extract it. We now determine the chemical potential $\mu (T)$ as
a function of absolute temperature $T$ \ for any positive space dimension, $%
d>0.$ This requires solving (\ref{FFD0})\ \ numerically when $d\neq 2$.
Whereas $\mu (T)$ in both $d=2,\ 3$ \ is well-known to \textit{decrease} in $%
T$ \ from the constant Fermi energy $E_{F}$ (which depends only on the
number density of fermions and on their mass) at $T=0$, towards the
well-known classical value diverging logarithmically to $-\infty $, for all $%
d<2$ we find \textit{novel, anomalous behavior} consisting in a curious
temperature \textit{non-monotonicity}: $\mu (T)$ first increases
quadratically as $T$ is increased, then changes curvature, acquires a
maximum, and finally decreases monotonically to the classical value. This
translates into a ``shoulder'' or ``hump'' in the heat capacity as function
of temperature, and may be relevant in the study of quantum ``dots,''
``wires'' and ``wells'' of modern-day opto-electronics \cite{Corcoran, corcoran1}.

Results are exhibited in Fig. 1 for several values of $d$. The unexpected 
\textit{rise} in $\mu (T/T_{F})$ for $d<2$ with increasing $T/T_{F}$ is
novel and anomalous, though it was reported in Ref. \cite{ariel} and more
completely in Ref. \cite{tsallis} albeit with some errors. For $d=1$ it is
graphed in Ref. \cite{kittel-kroemer} p. 192 for low temperatures, but
without comment. Using the large-$z$ expansion (Ref. \cite{Path} p. 510) $%
f_{\sigma }(z)\simeq (\ln z)^{\sigma }/\Gamma (\sigma +1)+\left( \pi
^{2}/6\right) (\sigma -1)\left( \ln z\right) ^{\sigma -2}/\Gamma (\sigma
)+\cdots $ the $d$-dimensional Sommerfeld expansion for $d\neq 2$ becomes 
\begin{equation}
\mu (T)/E_{F}\quad \smash
{\mathop{\relbar\joinrel\longrightarrow}\limits_{{T/T_F}\to 0}}\quad
1-(d/2-1)(\pi ^{2}/6)\left( T/T_{F}\right) ^{2}+O(T^{4}).
\end{equation}
Note that the first correction to unity is \textit{positive} for all $d\leq
2 $, and since for large enough $T$ the chemical potential must diverge
negatively to approach the classical value a ``hump'' will emerge. Using the
small-z expansion $f_{\sigma }(z)\simeq z-z^{2}/2^{\sigma }+\cdots $, one
gets 
\begin{equation}
\mu (T)/E_{F}\quad \smash
{\mathop{\relbar\joinrel\longrightarrow}\limits_{{T/T_F}\to \infty}}\
-\left( T/T_{F}\right) \ln \left[ \Gamma (d/s)\left( T/T_{F}\right)
^{d/2}(d/2)\right] ,
\end{equation}
or the well-known classical limit for large $T$.

From (\ref{TGP}) and (\ref{est}) we obtain this expression for the entropy $S
$ in dimensionless form 
\begin{equation}
S/Nk_{B}=({d/2}+1)\frac{f_{{d/2}+1}(z)}{f_{d/2}(z)}-d/2\frac{f_{d/2}(z)}{%
f_{d/2-1}(z)}\quad {\smash
{\mathop{\relbar\joinrel\longrightarrow}\limits_{T\to 0}}\quad }d\frac{{\pi }%
^{2}}{6}\left( \frac{k_{B}T}{E_{F}}\right) \quad {\smash
{\mathop{\relbar\joinrel\longrightarrow}\limits_{T\to 0}}\quad 0,}
\label{ent}
\end{equation}
where the $T\rightarrow 0$ limit is obtained by using the large-$z$
expansion for $f_{d/2}(z)$, and clearly complies with the third law of
thermodynamics.

\subsection{Internal energy}

The internal energy\ can be obtained from (see p. 159 of Ref. \cite{Path}) 
\begin{equation}
U(T,V)=-k_{B}T^{2}{\frac{\partial }{\partial T}}\left( {\frac{\Omega }{k_{B}T%
}}\right) _{V,{\Large z}}.  \label{u}
\end{equation}
Substituting (\ref{TGP}) in (\ref{u}) and comparing with (\ref{TGP}) we find
that (recalling that $V\equiv L^{d}$) 
\begin{equation}
U\left( T,V\right) =\frac{d}{2}\frac{A_{d}}{\beta ^{{d/2}+1}}f_{d/2+1}(z)=%
\frac{d}{2}\Omega =\frac{d}{2}PV.  \label{u3}
\end{equation}
Eq. (\ref{u3}) is a generalization of the relation $PV=\frac{2}{3}U$ for an
ideal gas of fermions (and in fact, also bosons), in the nonrelativistic
limit. Comparing (\ref{N}) with the last equation we find 
\begin{equation}
U\left( T,V\right) =\frac{d}{2}\frac{N}{\beta }\frac{f_{d/2+1}(z)}{f_{d/2}(z)%
}.  \label{U}
\end{equation}
For ${T\rightarrow 0,}$ using the large-$z$ expansion for $f_{d/2}(z)$\ one
has\ 
\begin{equation}
2U(T,V)/dNE_{F}\ {\smash
{\mathop{\relbar\joinrel\longrightarrow}\limits_{T\to 0}}}\left( \frac{2}{d+2%
}\right) +\frac{\pi ^{2}}{6}\left( \frac{T}{T_{F}}\right) ^{2}+\cdots .
\label{T0}
\end{equation}
In the classical limit\ ${T\rightarrow \ }{\infty }$, $z<<1,$ we again use$\
f_{\sigma }(z)\simeq z-z^{2}/2^{\sigma }+\cdots $ and from (\ref{U}) get 
\begin{equation}
U/N\smash {\ \mathop{\relbar\joinrel\longrightarrow}\limits_{T\to \infty}\ \
}\frac{d}{2}{k}_{B}T,  \label{Tinf}
\end{equation}
in accordance with the equipartition theorem in $d$ dimensions.

\subsection{Specific heat}

The specific heat at constant volume $C_{V}$ $\equiv \left[ {\partial }%
U(T,V)/\partial T\right] _{N,V}$ follows from (\ref{U}) 
\begin{equation}
C_{V}(V,T)=\frac{d}{2}A_{d}\left[ (d/2+1)k_{B}\frac{f_{d/2+1}(z)}{\beta ^{{%
d/2}}}+\frac{1}{z}\left( {\frac{\partial z}{\partial T}}\right) _{N,V}\frac{%
f_{d/2}(z)}{\beta ^{d/2+1}}\right] .  \label{cv}
\end{equation}
Substituting (\ref{N}) in the last equation, this becomes 
\begin{equation}
\frac{2C_{V}(V,T)}{dNk_{B}}=\left( d/2+1\right) \frac{f_{d/2+1}(z)}{%
f_{d/2}(z)}-\left( d/2\right) \frac{f_{d/2}(z)}{f_{d/2-1}(z)},
\end{equation}
where we used the relation 
\begin{equation}
\frac{1}{z}\left( {\frac{\partial z}{\partial T}}\right) _{N,V}=-k_{B}\beta
(d/2)\frac{f_{d/2}(z)}{f_{d/2-1}(z)},
\end{equation}
which is extracted from the (vanishing) derivative with respect to $T$ of
the number equation (\ref{N}).

Equations (\ref{T0}) and (\ref{Tinf}) for large-and small-$z$ may be
differentiated to yield the specific heat as\ 
\begin{equation}
C_{V}/Nk_{B}\smash {\ \mathop{\relbar\joinrel\longrightarrow}\limits_{T\to
0}\ \ }({d/2)}\frac{{\pi }^{2}}{3}\left( \frac{T}{T_{F}}\right)  \label{c0}
\end{equation}
and\ 
\begin{equation}
2C_{V}/dNk_{B}\smash {\ \mathop{\relbar\joinrel\longrightarrow}\limits_{T\to
\infty}\ \ }\ 1+\frac{z}{2^{d/2+1}}\left( 1-d/2\right) .  \label{cinf}
\end{equation}


Fig. 2 (top panel) illustrates the ``hump'' developed by the $C_{V}\ (T)$
for all $d<2$. These peculiar results for the ideal Fermi gas in $d<2$
dimensions, manifesting ``structure'' in the form of anomalous
(non-monotonic) behavior in the chemical potential and in the specific heat,
contrasts sharply with the ideal \textit{Bose }gas where the ``structure''
appears for all $d>2,$ (see Fig. 2.5 of Ref. \cite{Ziff} for integer $d,$
and Fig. 2 of Ref. \cite{Agui} for all $d>2$). The (sharp) structure observed
is the Bose-Einstein condensation (BEC) whose signature is a ``cusp'' in the
specific heat at the critical transition temperature for all $2<d\leq 4$,
and a ``jump'' in its value there for all $4<d<\infty $.

\section{Sound velocities}

To further exhibit the anomalous behavior for $d<2$ we have also calculated
mechanical properties, e.g., the isothermal $c_{T}$ and adiabatic $c_{S}$
velocity of sound in the IFG. These are defined as 
\begin{equation}
mc_{T}^{2}=\left( \frac{\partial P}{\partial n}\right) _{T}{,\ \qquad \ \ \
\ }mc_{{\LARGE S}}^{2}=\left( \frac{\partial P}{\partial n}\right) _{S},
\label{ct1}
\end{equation}
\bigskip where from (\ref{TGP}) the pressure $P\equiv -\Omega /V$ is given
by 
\begin{equation}
\ P=\left( A_{d}/L^{d}\right) \ \frac{f_{d/2+1}(z)}{\beta ^{{d/2}+1}}.
\label{pv}
\end{equation}

\subsection{Isothermal sound velocity}

From (\ref{ct1}) and (\ref{pv}) one has 
\begin{equation}
m{c}_{T}^{2}(T)=\frac{\left( A_{d}/L^{d}\right) }{\beta ^{{d/2}+1}}\left( 
\frac{\partial f_{{d/2}+1}(z)}{\partial n}\right) _{T}.  \label{velT}
\end{equation}
Using the relation 
\begin{equation}
\left( \frac{\partial f_{{d/2}+1}(z)}{\partial n}\right) _{T}=\beta f_{{d/2}%
}(z)\left( \frac{\partial \mu }{\partial n}\right) _{T}=\frac{1}{n}\frac{f_{{%
d/2}}(z)}{f_{{d/2-1}}(z)},  \label{rel}
\end{equation}
which is obtained from (\ref{N}), we find 
\begin{equation}
m{c}_{T}^{2}(T)=k_{B}T\frac{f_{{d/2}}(z)}{f_{{d/2-1}}(z)}.  \label{ct}
\end{equation}
Normalizing with $E_{F}\equiv mv_{F}^{2}/2=k_{B}T_{F}$ (\ref{ct}) we have
\begin{equation}
\lbrack {c}_{T}(T)/v_{F}]^{2}=\frac{1}{2}(T/T_{F})\frac{f_{{d/2}}(z)}{f_{{%
d/2-1}}(z)}.
\end{equation}
Using the large-$z$ expansion for $f_{d/2}(z)$ as $T\rightarrow 0$ we find 
\begin{equation}
\lbrack {c}_{T}(T)/v_{F}]^{2}\quad \smash
{\mathop{\relbar\joinrel\longrightarrow}\limits_{{T/T_F}\to 0}}\ \mu /\left(
dE_{F}\right) +\frac{\pi ^{2}}{12}\left( \frac{d}{2}-1\right) \frac{%
(T/T_{F})^{2}}{\mu /E_{F}}  \label{v0}
\end{equation}
or, since at $T/T_{F}=0,\mu =E_{F}$ one gets the familiar result
\begin{equation}
{c}_{T}(0)=v_{F}/\sqrt{d}.  \label{vt0}
\end{equation}
In the limit of small $z,$ $f_{\sigma }(z)$ $\rightarrow z$ as $T\rightarrow
\infty $ we have 
\begin{equation}
\lbrack {c}_{T}(T)/v_{F}]^{2}\ \smash
{\mathop{\relbar\joinrel\longrightarrow}\limits_{{T/T_F}\to \infty}}\
T/2T_{F},  \label{vinf}
\end{equation}
which corresponds to the classical limit. From (\ref{v0}) and (\ref{vinf}),
we conclude that ${c}_{T}(T)/v_{F}$ \textit{decrease} in $T$ \ for all $d<2$
from $1$ at $T=0,$ then changes curvature, acquires a minimum, and finally
increases linearly as $T$ to the classical value. In Fig. 3 we plot $%
c_{T}^{2}(T)/c_{T}^{2}(0)$ where $c_{T}(T)$ is the isothermal sound velocity
for an IFG as function of temperature $T/T_{F}$ for $d=1/4$, $1/2$, $1$, $2$
and $3$. Note that $c_{T}$ develops a ``trough'' for $d<2$ which deepens as $%
d$ decreases, more clearly seen in the bottom left panel where we show the
minimum values in the isothermal sound velocity compared with $c_{T}(0)$ as
function of Log $d$, and on the values of the temperature $T_{\min }$
corresponding to the minimum in top panel. Thus $c_{T}^{2}(T)$ never becomes
negative so that $c_{T}(T)$ is always real.

\subsection{Adiabatic sound velocity}

The adiabatic sound velocity $c_{S}$ in a Fermi gas is obtained from the
adiabatic state equation for an ideal gas\ in $d$-dimensions, namely (Ref. 
\cite{Path}, p. 229)

\begin{equation}
PV^{\gamma }= \hbox{const} \equiv N^{\gamma }{c},  \label{adia}
\end{equation}
where $\gamma \equiv 1+2/d,$ $N$ is the particle number and $c$ some
constant. Hence 
\begin{equation}
P=n^{\gamma }c.  \label{ps}
\end{equation}
Using (\ref{ps}) in the second equation of (\ref{ct1}) we find\ \ \ \ 
\begin{equation}
mc_{S}^{2}\ =\ \left( \frac{dP}{dn}\right) _{N,S}=\frac{\gamma P}{n}.
\label{cs}
\end{equation}
Substituting $\gamma $ and (\ref{ps}) in (\ref{cs}) we have
\begin{equation}
mc_{S}^{2}\ =\ \left( \frac{2}{d}+1\right) \frac{P}{n}=\left( \frac{2}{d}%
+1\right) k_{B}T\frac{f_{d/2+1}(z)}{f_{d/2}(z)},  \label{csq}
\end{equation}
from which is obviously monotonic for all $d$.

\section{Conclusions}

After constructing the grand potential, thermodynamic properties were
determined along with densities-of-states for an ideal Fermi gas in $d$
dimensions, integer or not. For $0<d<2$ dimensions these properties develop
a \textit{hump} in the chemical potential\ $\mu (T)$ and in the specific
heats $C_{V}(T)$ and a \textit{trough }in the isothermal velocities of sound 
$c_{T}(T)$. This \textit{structure} contrasts with the case of the ideal
Bose gas structure ocurring, however, for $d>2$ commonly associated with
BEC, and characterized by a \textit{cusp} in the specific heat when $2<d<4$
and a \textit{jump} for all $4<d<\infty $.

\section{Acknowledgments}

We thank Prof. V.V. Tolmachev for discussions and acknowledge partial
support from UNAM-DGAPA-PAPIIT (M\'{e}xico) \# IN106401 and CONACyT
(M\'{e}xico) \# 27828 E. M.G. thanks CONACyT for a scholarship.

\begin{figure}[tbh]
\centerline{\psfig{file=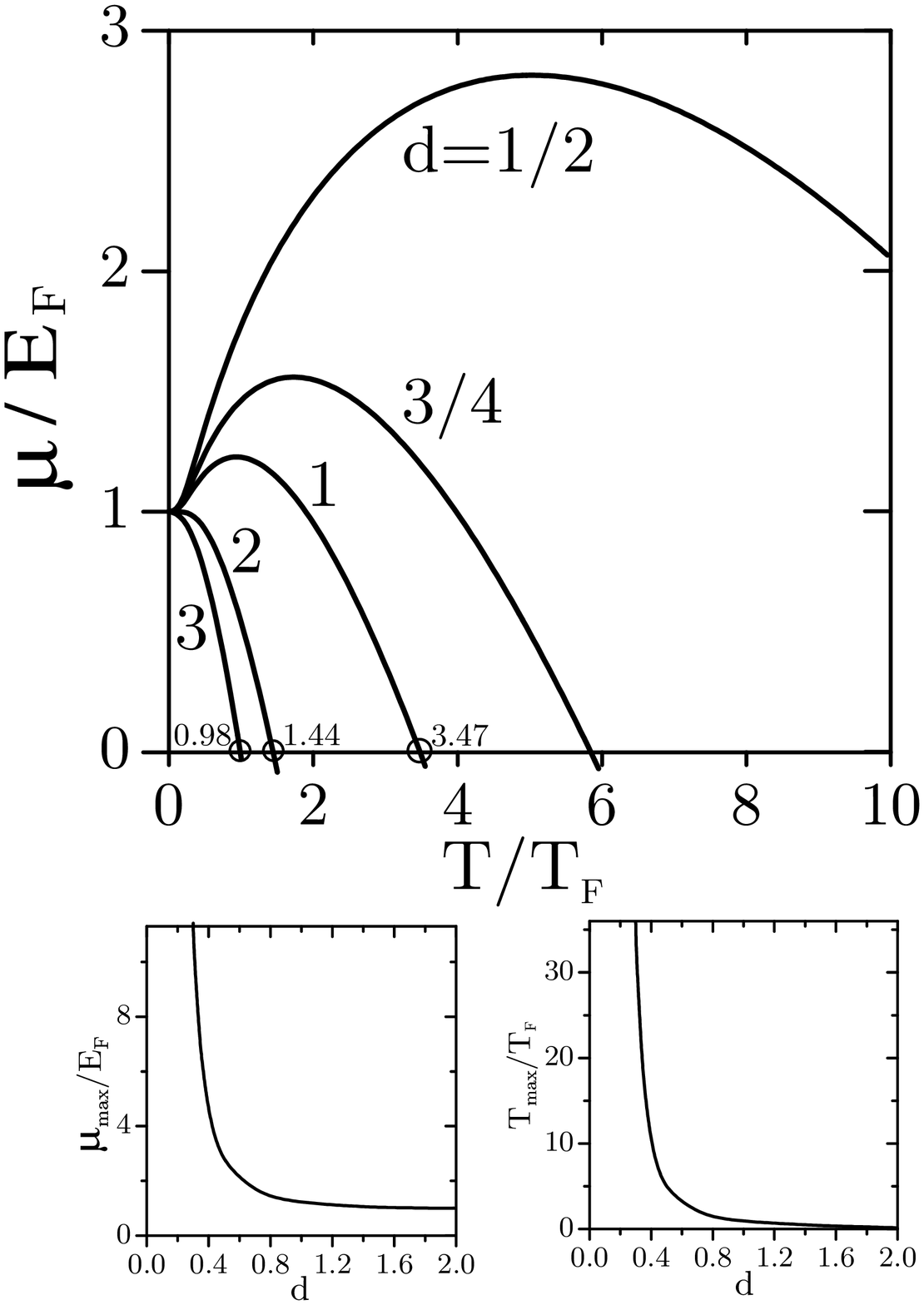,height=7.50in,width=6.0in}}
\caption{Chemical potential (in units of $E_{F}$) for an IFG in $d=1/2,$ $%
3/4, $ $1,$ $2$ and $3$ dimensions as functions of temperature $T$ (in units
of $T_{F}$). On the left bottom panel we show the maximum values in the
chemical potential as function of $d$, and on the right the values of the
temperature $T_{\max }$ corresponding to the maxima in top panel.
}
\end{figure}

\begin{figure}[tbh]
\centerline{\psfig{file=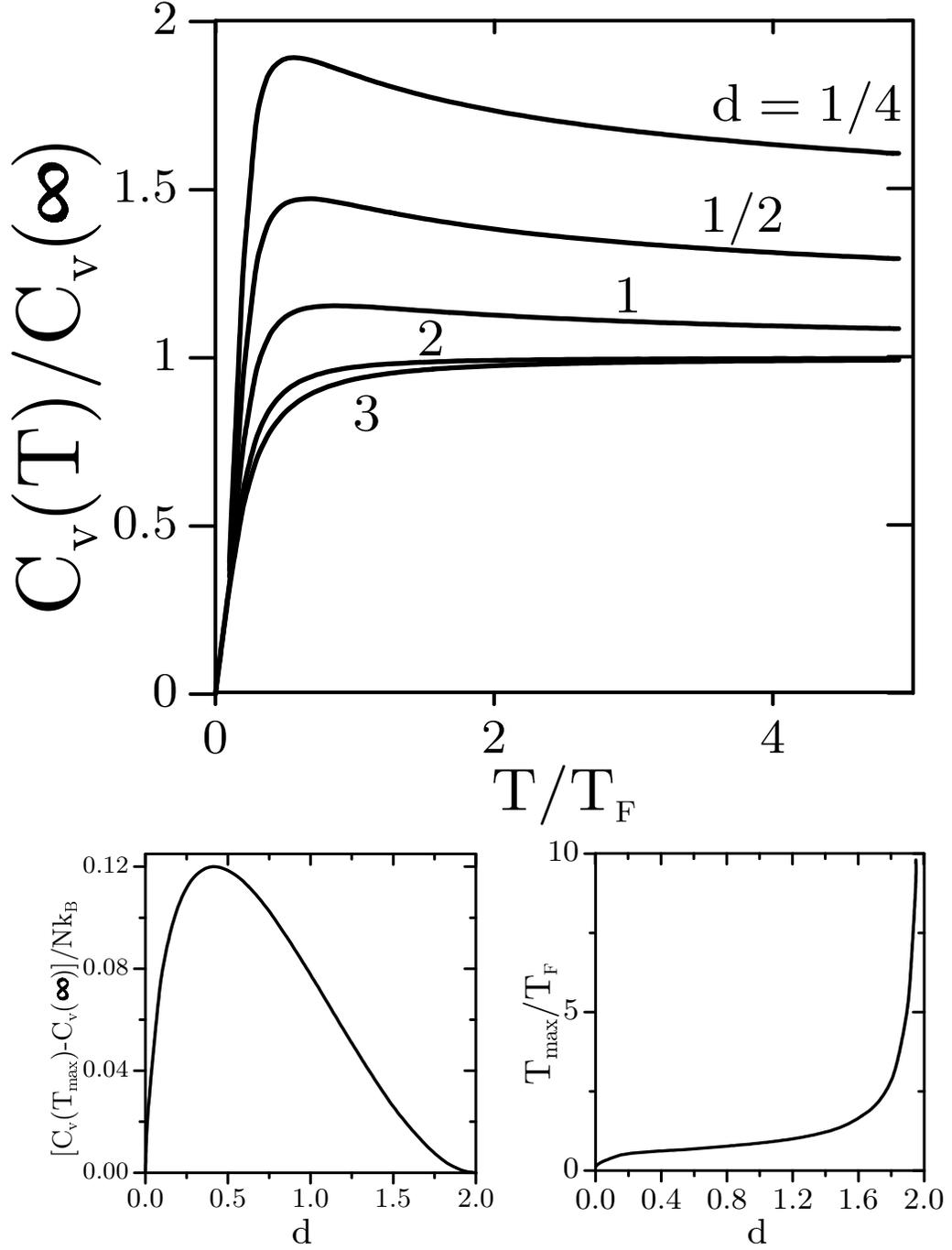,height=7.50in,width=6.0in}}
\caption{Specific heat for an IFG in $d=1/4,$ $1/2,$ $1,$ $2$ and $3$
dimensions. On the left bottom panel we show the maximum values of the
specific heat as function of $d$ and on the right the values of the
temperature $T_{\max }$ corresponding to the maxima in top panel.
}
\end{figure}

\begin{figure}[tbh]
\centerline{\psfig{file=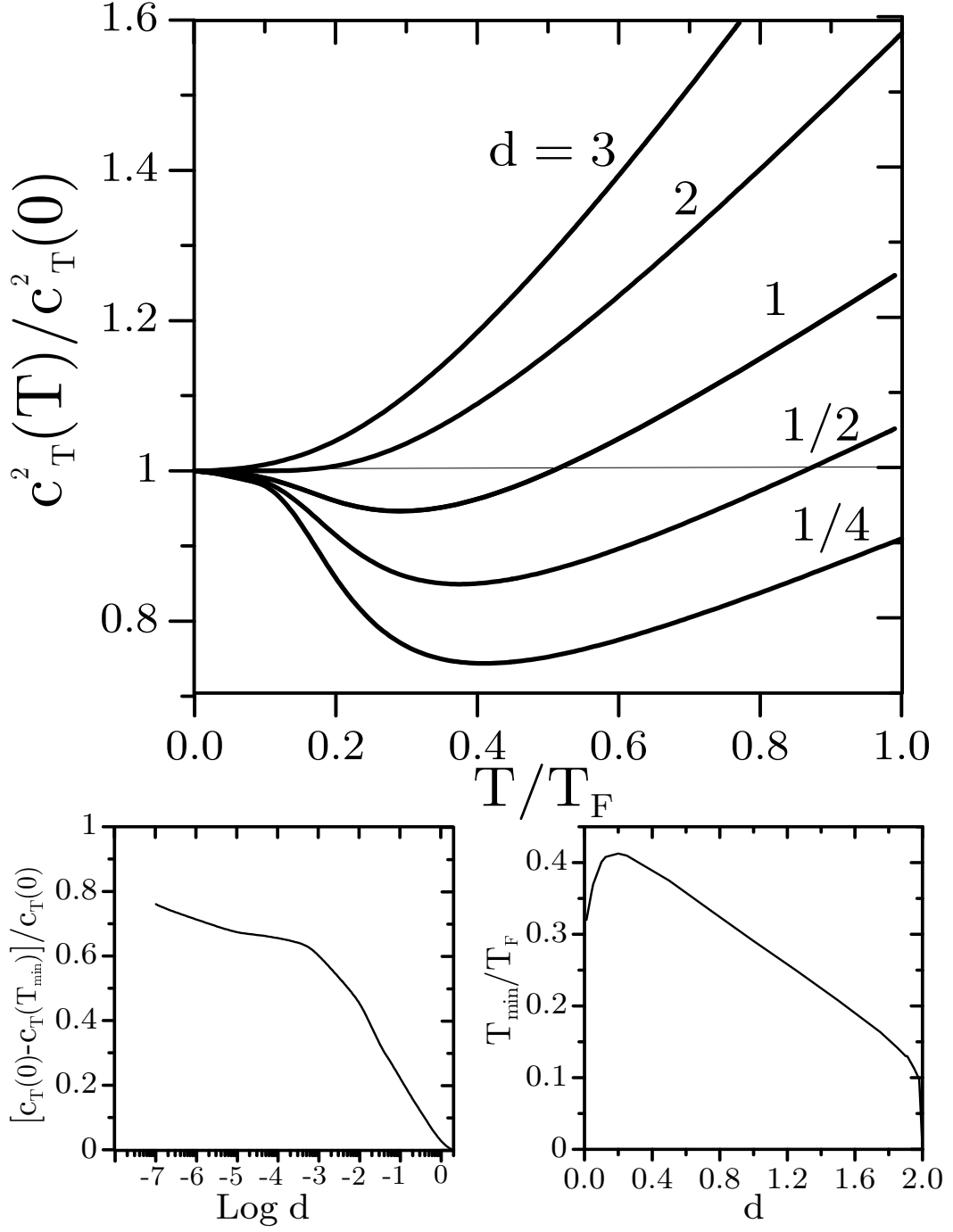,height=7.50in,width=6.0in}}
\caption{The quantity $c_{T}^{2}(T)/c_{T}^{2}(0)$ where $c_{T}(T)$ is the
isothermal sound velocity for an IFG as function of temperature $T/T_{F}.$
On the left bottom panel we show the minimum values in the isothermal sound
velocity compared with $c_{T}(0)$ as function of Log $d$, and on the values
of the temperature $T_{\min }$ corresponding to the minimum in top panel.
}
\end{figure}

\end{document}